\documentstyle[12pt,psfig]{article}

\def\thefootnote{\fnsymbol{footnote}}

\newcommand{\be}{\begin{equation}}	
\newcommand{\ee}{\end{equation}}	
\newcommand{\beq}{\begin{eqnarray}}	
\newcommand{\eeq}{\end{eqnarray}}	
\newcommand{\beqstar}{\begin{eqnarray*}}	
\newcommand{\eeqstar}{\end{eqnarray*}}	

\newcommand{\gsim}{ \mathop{}_{\textstyle \sim}^{\textstyle >} }
\newcommand{\lsim}{ \mathop{}_{\textstyle \sim}^{\textstyle <} }
\newcommand{\vev}[1]{ \left\langle {#1} \right\rangle }

\newcommand{\ov}[1]{\overline{#1}}

\newcommand{\upq}{U(1)_{\rm PQ}}
\newcommand{\fpq}{f_{\rm PQ}}

\begin{document}

\begin{titlepage}

\begin{minipage}[t]{3in}
\begin{flushleft}
\today
\end{flushleft}
\end{minipage}
\hfill
\begin{minipage}[t]{3in}
\begin{flushright}
EFI-2001-09\\
ANL-HEP-PR-01-021
\end{flushright}
\end{minipage}

\begin{center}
{\Large \bf Axions and a Gauged Peccei-Quinn Symmetry} 

\vskip .5in 

{\normalsize \bf Hsin-Chia Cheng$^1$,
 David Elazzar Kaplan$^{1,2}$

\vskip .5in
 
{\small \it $^1$Enrico Fermi Institute, The University of Chicago, Chicago, 
  IL 60637, USA\footnote{Email: hcheng@theory.uchicago.edu, 
  dkaplan@theory.uchicago.edu}\\
 $^2$High Energy Physics Division, Argonne National Laboratory,
  Argonne, IL 60439, USA}   
     }

\vskip .5in

\begin{abstract}
{\normalsize  
The axion solution to the strong CP problem requires an anomalous global
$U(1)$ symmetry.  We show that the existence of such a symmetry is a natural
consequence of an extra dimension in which a gauged $U(1)$ is spontaneously
broken on one of two branes, leaving an accidental global symmetry on the
other brane.  Depending on where the standard model matter lives, the
resulting axion can be either the DFSZ or hadronic type.  Gaugino-mediated
supersymmetry breaking fits comfortably in our framework.  In addition, 
we present a model in which the supersymmetry-breaking and Peccei-Quinn 
breaking scales are naturally of the same size.}

\end{abstract}

\vskip 1in

\end{center}
\end{titlepage}

\renewcommand{\thefootnote}{\arabic{footnote}}
\setcounter{footnote}{0}
\pagestyle{plain} 

\section{Introduction}

The Peccei-Quinn mechanism~\cite{Peccei:1977hh} is 
probably the most
elegant solution to the strong CP problem. It introduces a new global
$\upq$ symmetry, spontaneously broken at a scale $\fpq$. The CP violating
parameter $\ov{\theta}$ in the QCD Lagrangian,
\be
\ov{\theta}\, \frac{g_3^2}{32\pi^2}\, G_a^{\mu\nu} \tilde{G}_{a\mu\nu},
\ee
then becomes a dynamical degree of freedom, corresponding to the 
Nambu-Goldstone boson of the broken $\upq$ symmetry, called the axion,
$\ov{\theta}\to a(x)/\fpq$~\cite{Wilczek:1978pj}. 
The $\upq$ is anomalous with respect to
$SU(3)_C$, and the QCD instanton effects generate a periodic potential
for the axion,
\be
V(\ov{\theta}) \sim m_{\pi}^2 f_{\pi}^2 (1-\cos\ov{\theta}).
\ee
The minimum is at the CP conserving value $\ov{\theta}=0$, therefore
it solves the strong CP problem.

Astrophysics and cosmology put strong bounds on the $\upq$ breaking
scale, 
\be
10^9\, {\rm GeV}\, \lsim \, \fpq\, \lsim \, 10^{12}\, {\rm GeV}.
\ee
The lower bound comes from the energy losses in globular-cluster stars
and the supernova SN~1987A~\cite{Raffelt:1999tx}, 
and the upper bound comes
from the requirement that the axion density does not overclose the 
universe~\cite{Preskill:1983cy,Davis:1986xc,Harari:1987ht}. 
If there is some late entropy production  which dilutes 
the axion density, the upper bound may be raised to 
$\sim 10^{15}$~GeV~\cite{Steinhardt:1983ia}. 
The bounds also depends on the nature of the axion,
which can be divided into two classes, the DFSZ axion~\cite{Zhitnitsky:1980he}
 and
the hadronic (or KSVZ) axion~\cite{Kim:1979if}. In the DFSZ-type model, 
it requires two Higgs doublets and the standard
model quarks and leptons carry $\upq$ charges, while in the hadronic axion
model, one introduces new heavy quarks which carry Peccei-Quinn charges
and the ordinary quarks and leptons are neutral under $\upq$.
For the hadronic axion, if the axion-photon coupling 
is small, ($C_{a\gamma\gamma}<0.1$),
there is a small window at $\fpq \sim 10^6$ GeV
which may be allowed~\cite{Chang:1993gm,Moroi:1998qs}.

The problem of the Peccei-Quinn mechanism is to understand why there
exists such a global symmetry. Global symmetries are argued to be broken
by quantum gravity effects~\cite{Giddings:1988cx}.
In fact, even without considering quantum gravity, $\upq$ is not an
exact symmetry, explicitly broken by the anomaly from QCD. If it is
an accidental symmetry at the renormalizable level, one expects that
it will be violated by the Planck scale physics, which may be represented
by the Planck scale suppressed operators. However, such explicit
$\upq$ violating effects contribute to the axion potential and can
change the minimum, therefore render the strong CP solution unnatural.
It has been shown that the higher dimensional operators have to be suppressed
by more than $M_{\rm pl}^8$ in order to satisfy the neutron electric
dipole moment constraint~\cite{Kamionkowski:1992mf,Barr:1992qq,Holman:1992us}.
There have been attempts
which use additional discrete or gauge symmetries to forbid these
Peccei-Quinn violating interactions to 
such high orders~\cite{Holman:1992us,Chun:1992bn}.

In this Letter we show that this problem may be easily solved with
a {\it gauged} $\upq$ if the gauge fields propagate in an extra dimension. 
Extra dimensions provide a natural framework for the accidental 
global symmetries~\cite{Arkani-Hamed:1998sj,Cheng:1999fw,Arkani-Hamed:2001xj}.
If there are fields which are charged under some gauge symmetry localized
at different branes in the extra dimensions, then effectively there will 
be a separate symmetry on each brane if there are no light bulk fields
connecting them. If the gauge symmetry is broken by the fields on
two different branes, there will be two (sets of) Nambu-Goldstone bosons.
One linear combination is eaten by the gauge field and becomes heavy. 
The other corresponds to the Nambu-Goldstone boson of the broken 
accidental global symmetry. It can obtain some small mass if the accidental
symmetry is anomalous or broken by some bulk-brane interactions.
However, there will be no Planck scale physics violating the symmetry
on the branes since the symmetry is gauged.  The contributions from
the bulk fields will be exponentially suppressed if the bulk fields
are heavier than the inverse of the distance of the branes.

\section{Models}

We consider an extension of the supersymmetric (SUSY) standard model (SM)
with an additional $\upq$ gauge symmetry. All gauge fields propagate in 
an extra dimension.  We assume that the size of the extra dimension
is larger than the 4-dimensional Planck length, but smaller than
$\fpq^{-1}$. The SM matter fields, including  
right-handed
neutrinos, $Q_i,\, U^c_i,\, D^c_i,\, L_i,\, E^c_i, {\nu_R}_i,\, i=1,2,3$, 
and the two Higgs doublets, $H_U,\, H_D$, are localized on a 3-brane
(SM brane).
All the ordinary matter fields have charge $+1$ under $\upq$ and the 
Higgs fields have charge $-2$ so that the Yukawa couplings are allowed.
In addition, on the same brane there are $\upq$ charged SM 
singlets which are responsible for breaking the $\upq$ symmetry. 
We choose them to be $P(+2)$ and $N(-2)$ for simplicity.
When they get nonzero vacuum expectation values (VEVs) 
at the intermediate scale,
they can generate the Majorana masses for the right-handed neutrinos
through the $N {\nu_R}_i {\nu_R}_j$ interactions, and the $\mu$-term
through the non-renormalizable interactions 
\be
\frac{1}{M}\, P^2\, H_U\, H_D,
\label{mu}
\ee
where $M$ is the fundamental Planck scale.

To cancel the anomalies of the $\upq$, we need to add additional fields
charged under the SM gauge group and $\upq$. We assume that
these fields are localized on a different brane (hidden brane) 
from where the ordinary
matter resides. A simple choice consists of 3 pairs of $\ov{d},d(-2)$, 2 
pairs of $\ov{\ell},
\ell(-2)$ fields, which are vector-like under the SM gauge
group and transform like the $SU(2)_W$-singlet down-type quarks and the 
$SU(2)$-doublet leptons, and 3 SM gauge singlets $X_i(+4),\,
i=1,2,3$.\footnote{This choice is not unique. Discussion of finding
anomaly-free combinations can be found in Refs.~\cite{Cheng:1999hc}.}
We require that one of the $X_i$ fields 
(denoted by $X$) gets a nonzero VEV and gives large masses to the 
$\ov{d},d$ and $\ov{\ell}, \ell$ fields through the interactions,
$X\ov{d}d,\, X\ov{\ell}\ell$.
The total particle content is anomaly free which can be seen
easily by noticing that $\upq$ can be embedded into the $E_6$ gauge
group. To cancel the anomalies everywhere there should be Chern-Simons
terms in the bulk which provide the anomaly inflow from one brane to 
another~\cite{Callan:1985sa}. 

Assuming that there is no light $\upq$ charged field 
in the bulk which interact with fields on both branes, there is effectively
one (anomalous) $\upq$ symmetry on each brane. After they are broken
by the VEVs of the $P,\,N$ and $X$ fields on these two branes, there
are two corresponding Nambu-Goldstone bosons. One linear combination
is eaten by the $\upq$ gauge field and becomes heavy. The other remains
light and gets a small mass from the anomaly, due to the Chern-Simons
terms in the bulk. This becomes the axion
which relaxes $\ov{\theta}$ to zero. There is no Planck scale physics
violating $\upq$ because it is a gauge symmetry. If there are heavy
bulk fields charged under $\upq$ which couple to both branes, they can 
contribute to the axion potential because they connect the two $\upq$
symmetries on the two branes. However, their effects are suppressed
exponentially and can be made safe easily if their masses are much
larger than the inverse of the distance, $L$, of the two branes.

The axion in this model can be either the DFSZ type or the hadronic
type depending on the $\upq$ breaking scales on the two branes.
If the $X$ VEV is larger than the VEVs of $P$ and $N$, the axion lies
mostly on the SM brane, and it is the DFSZ axion. 
On the other hand, if the $X$ VEV
is smaller, the axion lies in the hidden brane, it becomes the 
hadronic axion. It interpolates between the two type of axions if the
VEVs on the two branes are comparable.

This two-brane setup also fits well with the gaugino-mediated 
SUSY breaking scenario~\cite{Kaplan:2000ac}. If supersymmetry
is broken on the hidden brane, SUSY breaking
can be transmitted to the SM sector through the gauge
fields in the bulk. It solves the supersymmetric flavor problem
by giving flavor-universal contributions to the superpartners of the
SM quarks and leptons. The coincidence of the SUSY
breaking scale and the Peccei-Quinn breaking scale also hints at the 
tantalizing possibility that these two breaking scales are correlated.
In the following we describe two models which give rise to a a hadronic 
axion and the DFSZ axion respectively.

\subsection{A hadronic axion model}

We assume that SUSY breaking occurs on the hidden brane which contains
the $\ov{d},\, d,\, \ov{\ell}, \, \ell$, and the $X$ fields, 
so these fields and
the bulk gauge fields can couple directly to the SUSY-breaking
field and obtain SUSY-breaking masses of the order of the weak scale.
The ordinary squarks and sleptons receive SUSY-breaking masses from
the running contributions of the gaugino masses below the scale $L^{-1}$. 
These contributions
are positive and comparable to the gaugino masses due to the large logarithm
enhancement. The $\upq$ gauge symmetry can be broken on the SM brane if
$P$ and $N$ fields have negative squared masses. This happens if
the coupling between the $\upq$ gauge field and the
SUSY-breaking field is suppressed so that the $\upq$ SUSY-breaking
gaugino mass is vanishingly small. Then, the dominant contributions to
the SUSY-breaking masses of the SM singlets
on the SM brane ($P,\, N,\, \nu_R$) are the two-loop running 
contributions of the soft
masses of the hidden brane fields, and the 
anomaly-mediated contribution~\cite{Randall:1999uk}.
These contributions to the squared masses of the scalars are negative.
The fields $P$, $N$ can then get large VEVs to break $\upq$. The 
right-handed sneutrinos are prevented from getting VEVS due to the
interactions $N {\nu_R}_i {\nu_R}_j$. 

The VEVs of the $P$, $N$ fields
can be stabilized by the non-renormalizable interactions,\footnote{
The mass term $PN$ may be forbidden by a parity under which $P$ 
changes sign. Alternatively, we can assign $P$ with a different charge
and cancel the anomaly with fields on the hidden brane. For $P$ field
of charge $+2p, \, (p>2)$, The VEVs can be stabilized by the superpotential
$P\,N^p\,/M^{p-2}$ and the $\mu$-term can be generated by $P\, N^{p-2}\,
H_U\, H_D\,/M^{p-2}$. The SM brane sector then resembles the model of 
Ref.~\cite{Murayama:1992dj}.}
\be
\frac{\lambda}{M}\, P^2\, N^2,
\ee
and will be of the order
\be
v\, \sim\, \sqrt{\frac{m_{\tilde{P}} M}{\lambda}}\, 
\sim\, \frac{10^9\, \rm{GeV}}{\sqrt{\lambda}},
\ee
where $m_{\tilde{P}}$ is the size of the soft scalar SUSY-breaking mass of 
the $P$ 
field, and is expected to be ${\cal O}$(1 GeV), and $M$ is the fundamental
Planck scale
which is close to but somewhat smaller than the effective
four-dimensional Planck scale $2.4\times 10^{18}$ GeV because of the 
existence of extra dimensions larger than the fundamental length scale.
We assume that the coupling $\lambda$ is small ($\lambda < 10^{-2}$,
$v > 10^{10}$ GeV) so that the $\mu$-term of the Higgs superpotential
can be generated by the operator
\be
\frac{\lambda'}{M}\, P^2\, H_U\, H_D.
\ee

On the hidden brane the Peccei-Quinn symmetry can be broken radiatively.
The interactions
\be
\kappa_d\, X\,\ov{d}\, d, \,\,\, \kappa_{\ell}\, X\, \ov{\ell}\, \ell,
\ee
can drive the SUSY-breaking squared mass of the $X$ scalar, $m_{\tilde{X}}^2$ 
to negative in running down to low energies if $\kappa_d$, $\kappa_{\ell}$
are big enough. Including radiative corrections, 
the size of the $X$ VEV will be stabilized at the scale where $m_{\tilde{X}}^2$
changes sign~\cite{Coleman:1973jx}, which depends on the couplings $\kappa_d$,
$\kappa_{\ell}$ and the soft masses of the $X$, $\ov{d}$, $d$, $\ov{\ell}$,
$\ell$ fields. If the VEV of the $X$ field, $v_X$, is smaller than the 
VEVs of the $P$, $N$ fields, $\upq$ gauge symmetry is mostly broken by the 
$P$, $N$ VEVs. The $X$ VEV then breaks the left over accidental global
Peccei-Quinn symmetry on the hidden brane and hence $\fpq \approx v_X$.
The resulting axion is of the hadronic type and lies mostly in the $X$ field.
The axion-photon coupling~\cite{Kaplan:1985dv}
\be
C_{a\gamma\gamma}= \frac{E_{\rm PQ}}{N_{\rm PQ}}-1.92 \pm 0.08,
\ee
is small in this model because the ratio of the electromagnetic and 
the color anomalies of Peccei-Quinn symmetry in the hidden sector,
${E_{\rm PQ}}/{N_{\rm PQ}}$, is 2.
Therefore, this model is viable if $v_X$ lies in the hadronic axion
window $\sim 10^6$ GeV or in the conventional 
range $10^{9}-10^{12}$ GeV (with $v_X<v$).

If $v_X$ is larger than $\sqrt{m_{\tilde{P}} M/\lambda}$, then the $N$
field will get a VEV of the order of $v_X$ while the $P$ field will
be prevented from getting a VEV due to the $\upq$ $D$-term interactions.
In this case, the axion will be a comparable mixture of the $X$ and $N$
fields with $\fpq \sim v_X$. However, we will need some other way to
generate the $\mu$-term because $\vev{P}=0$.

\subsection{A DFSZ axion model with correlated SUSY and Peccei-Quinn
breaking scales}

Now we include an explicit model of supersymmetry breaking using the
shining method~\cite{Arkani-Hamed:2001pv}.  This can be accomplished by
adding a pair of uncharged chiral superfields $\Phi,\Phi^c$ with mass $m$
to the bulk.  In the language of four-dimensional ${\cal N}=1$ superspace,
the superpotential now contains
\begin{equation}
\Phi^c(x_5) \left( \partial_5 + m \right) \Phi (x_5).
\end{equation}
Adding the source $- J \Phi^c \delta (x_5)$ on the hidden brane (at $x_5 =0$)
and a coupling $S \Phi \delta (x_5 - L)$ to a field $S$ on the SM brane
(at $x_5=L$) gives the following $F$-term equations:
\begin{eqnarray}
- F_{\phi^c}^* &=& (\partial_5 + m)\phi - J \delta (x_5),\nonumber\\
- F_{\phi}^* &=& (- \partial_5 + m)\phi^c + S \delta (x_5 - L),\nonumber\\
- F_S^* &=& \phi (x_5 = L).
\end{eqnarray}
The first and third lines cannot be made to vanish simultaneously.  
The first line vanishes if
\begin{equation}
\phi = \frac{J e^{-m x_5}}{1 - e^{-2 m L}},
\end{equation}
where we have assumed a compactification length of $2L$.  
The above gives $F_S\sim J e^{-m L}$ and thus supersymmetry is broken 
\cite{Arkani-Hamed:2001pv}.

The gauged $\upq$ symmetry is assumed to be broken on the hidden brane 
at a high scale ($> \sqrt{F_S}$). We need to introduce an additional
pair of $X_4(+4),\, \ov{X}(-4)$ fields on the hidden brane so that
there is a $D$-flat direction where $X,\,\ov{X}$ can get large VEVs.
The gauge $\upq$ breaking can occur
radiatively as described in the previous scenario, or by the superpotential 
interaction,
\be
Z(X\ov{X}- v_X^2).
\ee
This breaking will add
$D$-term contributions to soft masses if the soft masses $m_{\tilde{X}}^2$
and $m_{\tilde{\bar{X}}}^2$ are not equal.  The matter contributions will be 
universal, and are positive if $m_{\tilde{X}}^2 < m_{\tilde{\bar{X}}}^2$
(with negative contributions to Higgs soft masses).

To break the remaining anomalous global $\upq$ symmetry on the SM brane, we 
add a singlet $T$ to the SM brane with superpotential couplings
\begin{equation}
T (P N - k \Phi^c ).
\end{equation}
We see that $F_T = 0$ requires (assuming approximately equal soft masses)
\begin{equation}
P \sim N \sim \sqrt{k}\, e^{- m L/2}.
\end{equation}
If all couplings are of order unity in units of the fundamental 
Planck scale, a compactification length of $m L\sim 32$ gives a 
supersymmetry-breaking and the global $\upq$-breaking scale of 
a few times 10$^{10}$ GeV\footnote{Bounds on operators which explicitly
violate the global $\upq$ require any charged bulk field to have a mass
$m_c$ such that $m_c L \gsim 130$, 
or $m_c \gsim 4 m$~\cite{Kamionkowski:1992mf}.}.  A right-handed 
neutrino mass will be generated at the same scale, and a $\mu$ term of 
the correct size will be produced by the operator in equation (\ref{mu}).

In this simple model supersymmetry is broken on the SM brane. 
The scalar superpartners of the SM fermions can receive soft masses
via the contact terms which can be
flavor violating. Models with SUSY breaking on the hidden brane
so that gaugino-mediation is responsible for scalar masses 
can also be constructed
by including shining in both directions with sources on both the SM brane
and the hidden brane~\cite{Schmaltz:2000gy}. 

\section{Discussion and Conclusions}

A nice feature of the models discussed in the previous sections is that
$R$ parity is automatically conserved.  The $\upq$ gauge symmetry is broken
only by fields with even charges, while all the ordinary matter
superfields $Q_i,\, U^c_i,\, D^c_i,\, L_i,\, E^c_i, {\nu_R}_i$ have
charge +1. A $Z_2$ matter parity, equivalent to the
$R$ parity, is left unbroken, so the $R$ parity conservation is an
automatic consequence of the $\upq$ gauge symmetry.

There are several dark matter candidates in these type of the theories.
Axions with $\fpq \sim 10^{12}$ GeV have been known as a
popular cold dark matter candidate~\cite{Preskill:1983cy}.
The hadronic axion in the hadronic axion window, $\fpq \sim 10^6$ GeV,
can serve as a hot dark matter component~\cite{Moroi:1998qs}.
The superpartner of the axion, the axino, is also a good cold dark
matter candidate if it is the lightest supersymmetric
particle (LSP)~\cite{Covi:1999ty}, and it can relax the bounds on the
ordinary superpartner masses from the constraint $\Omega_{\chi} h^2 <1$
in the neutralino ($\chi^0$) LSP scenario. Finally, the superheavy
fields $\ov{d},\, d,\, \ov{\ell},\, \ell$ may also be the dark 
matter~\cite{Chung:1999zb}.

The low energy theory is simply the supersymmetric standard model (SSM)
with one or more SM singlet fields which contain the axion. Their
couplings to the SSM fields are highly suppressed by the intermediate
scale $\fpq$, so they are difficult to produce at the colliders.
However, if the axino is the LSP, the next-to-lightest supersymmetric
particle will decay to the axino. Although the average proper decay length
is likely much larger than a typical collider detector, with a large
sample of SUSY events the decay will occasionally happen in the detector,
giving rise to a spectacular signal~\cite{Martin:2000eq}. In addition,
if the lightest ordinary superpartner is a charged slepton, as can happen
in the gaugino mediation models~\cite{Kaplan:2000ac}, 
the slowly moving long-lived 
charged sleptons will produce highly ionizing tracks which will easily
be discovered~\cite{Feng:1998zr}.

In conclusion, we have shown that an accidental Peccei-Quinn global
symmetry can arise naturally in a theory with gauge fields propagating
in an extra dimension. The resulting axion from the broken accidental
Peccei-Quinn symmetry only receives its mass from the anomaly, 
but not from any Planck scale physics providing there are no other
$\upq$ charged bulk fields communicating between the two branes.
Therefore it provides a viable solution to the strong CP problem.
The similar setup can also be used to generate other possible global
symmetries and pseudo-Nambu-Goldstone bosons.
For example, the quintessence field which explains the dark energy
in the universe could be the ultra-light pseudo-Nambu-Goldstone boson 
from a broken accidental global symmetry in theories with
extra dimensions~\cite{Hill:1989vm,Carroll:1998zi}.


{\bf Acknowledgements} 
We would like to thank Chris Hill and Martin Schmaltz for discussion.
H.-C. Cheng is supported by the Robert R. McCormick Fellowship and by
DOE grant DE-FG02-90ER-40560.  D.~E.~Kaplan is supported by DOE grants
DE-FG02-90ER-40560 and W-31-109-ENG-38.

%
%
\newcommand{\Journal}[4]{{#1} {\bf #2} {(#3)} {#4}}
\newcommand{\APJ}{Ap. J.}
\newcommand{\CJP}{Can. J. Phys.}
\newcommand{\NC}{Nuovo Cimento}
\newcommand{\NP}{Nucl. Phys.}
\newcommand{\MPL}{Mod. Phys. Lett.}
\newcommand{\PL}{Phys. Lett.}
\newcommand{\PR}{Phys. Rev.}
\newcommand{\PRep}{Phys. Rep.}
\newcommand{\PRL}{Phys. Rev. Lett.}
\newcommand{\PTP}{Prog. Theor. Phys.}
\newcommand{\SJNP}{Sov. J. Nucl. Phys.}
\newcommand{\ZP}{Z. Phys.}


\end{document}